\documentstyle[aps,prb,preprint]{revtex}
\begin{document}
\draft
\title{Block-Spin Approach to Electron Correlations}
\author{P. Monthoux\cite{Monthoux}}
\address{National High Magnetic Field Laboratory and Department of Physics
\\Florida State University, Tallahassee, Florida 32306}
\date{\today}
\maketitle

\begin{abstract}

We consider an expansion of the ground state wavefunction of quantum
lattice many-body systems in a basis whose states are tensor products
of block-spin wavefunctions. We demonstrate by applying the method to
the antiferromagnetic spin-1/2 Heisenberg chain that by selecting the
most important many-body states the technique affords a severe
truncation of the Hilbert space while maintaining high 
accuracy.

\end{abstract}


\narrowtext

\section{Introduction}

The basic physical idea underlying the Renormalization Group (RG) approach
to critical phenomena and quantum field theory is that the many length
or energy scales are locally coupled. There is then a cascade 
effect in the whole system: the eV energy scales influence the 100 meV
energy scales. The 100 meV scales influence the 10 meV energy scales, etc.
The RG formalism can then be set up to reduce the problem into one
involving a finite number of degrees of freedom at each 
iterations\cite{Wilson}.

Wilson's solution of the Kondo problem\cite{Wilson} by means of RG
techniques inspired many attempts at a RG approach to quantum
lattice models\cite{RSRG}. These latter real space RG schemes employed a 
Kadanoff block-spin construction. The lattice is partitioned into blocks
of a few sites each and the Hamiltonian on a block is diagonalized. 
One seeks an effective Hamiltonian that describes the physics of the
low energy degrees of freedom in each block. The contribution of the
high energy states of the block is either ignored altogether
in the simplest approximation or incorporated by low order perturbation
theory. While generally good results were obtained for the ground state
state energy of the system, the ground state wavefunction was not
satisfactorily described. These simple effective Hamiltonian approaches
suffer from small energy denominators, e.g. the denominators involving
differences of the highest energy kept minus the lowest energy 
eliminated\cite{Glazek}.

The Density Matrix Renormalization Group (DMRG) algorithm\cite{DMRG}
solved the problems that plagued the simple real-space
RG schemes with two clever ideas. Instead of doubling the system
size at each RG iteration, one adds one site to the system
at each iteration. The wavefunctions of the N-site and N+1-site systems 
should be "close" to each other, thus allowing a truncation of the Hilbert
space without much loss in accuracy. The second idea is to use the
eigenstates of the block density matrix as a basis instead of the
eigenstates of the Hamiltonian. In two insightful papers, Rommer
and Ostlund\cite{Rommer} showed that in the case where the DMRG converges
to a fixed point, the quantum states in the thermodynamic limit with
periodic boundary conditions can be simply represented by a "matrix product
ground state". These states can be obtained through a simple variational
ansatz, in the construction of which the DMRG plays no essential role, 
aside from providing a guide to which representations to keep. Such
"matrix product states" are in fact exact ground states for some 
particular 1-d models\cite{Affleck,Acardi,Fannes,Klumper}. 
One can almost certainly attribute the success of the DMRG
method in 1-d to the clever choice of a correlated basis that
describes the physics of such systems so well.

By the same token, one can speculate that the difficulties encountered
in tackling 2-d problems with the DMRG method may be due to the fact that 
the "matrix product states" are not so well tuned to the physics of 2-d
systems, which is quite different from that of their 1-d counterparts.
It is usually advisable to use a symmetry-adapted basis with as many
symmetries of the eigenstates one is interested in as possible. But
any such matrix product state in 2-d cannot be translation invariant
and does not transform according to any of the irreducible
representations of the symmetry group of the lattice. This must
limit the accuracy achievable in 2-d with a matrix product ansatz
for the ground state. In the present work, a different correlated 
basis is used to expand the many-body wavefunction that can be used 
(in principle) in any dimension.

The main difficulty in carrying out the RG program for a quantum many-body
Hamiltonian is to find a representation of the problem (basis) for which
the assumption of locally coupled energy scales is valid, at least to 
a good approximation. We have carried out simple numerical experiments on
random matrices\cite{Random} and we found that for the many energy scales to be
locally coupled, the Hamiltonian matrix elements $\rm H_{i,j}$ have to
decay away from the diagonal. If the decay follows a power law 
(say $\rm H_{i,j}$ goes like $\rm 1/|E_i - E_j|$), the weights of
the basis states in the ground state wavefunction decay (on average) as the
inverse square of their energy. If the decay of the matrix elements away from
the diagonal is exponential, then the weights of the basis states in the 
ground state wavefunction decay (on average) 
exponentially with energy. In either case, one has a well
defined hierarchy of basis states, and one can envisage 
solving the problem by means of RG techniques. A renormalization procedure
for Hamiltonians such as those in light-front field theory that achieves this
local coupling of energy scales while avoiding the small denominator problem 
is presented in Ref.~\protect\cite{Glazek}. The bare Hamiltonian is transformed
by a particular similarity transformation. This transformation is such that
all matrix elements vanish which would have otherwise caused transitions 
with big energy jumps.

If one diagonalizes a strongly correlated system, such as the 24-site 
spin-1/2 Heisenberg chain in an uncorrelated basis (tensor product of
single site states), one finds that at energies five or six times the
exchange constant above the mean-field (or N\'eel) state, the weights
of the basis states of equal energy span six to eight orders of 
magnitude! Such a basis is useless for carrying out the RG program.

A suitable basis must therefore include some of the correlations present 
in the true many-body ground state. It must remain simple enough, however, 
so that the evaluation of matrix elements of the Hamiltonian can be done 
in a resonable amount of computer time, and thus a good compromise must
be found. In this paper, we examine a block-spin basis for expanding the
wavefunction of quantum lattice models with short-range interactions. 
Since the energy essentially depends on correlations on a scale
of the range of the interactions, it is dominated by short-range 
correlations. Such correlations can be taken into account by grouping
the sites into blocks of a few sites each and treating the interactions
among the sites in a given block non-perturbatively. There are of course 
other ways of taking into account short-range correlations, but the advantage 
of the block-spin construction is that is provides an orthonormal basis
and the evaluation of matrix elements of the Hamiltonian is straighforward.
The long-range correlations, i.e. correlations between sites in 
different blocks, can be taken into account by using a linear superposition
of tensor products of block-spin states. For large systems, 
it is impossible in practice to retain all such tensor products,
and one must truncate the expansion to the most important many-body
configurations.  We thus make use of the adaptive state selection
techniques used in the context of Quantum Chemistry\cite{StateSelection}. 
We demonstrate the method by applying it to a test problem whose exact
solution is known, the 40-site antiferromagnetic spin-1/2
Heisenberg chain. We find that it is possible
to recover 99.8\% of the correlation energy and to describe
the spin correlations at a distance of 20 lattice spacings with
an accuracy of 1.5\% with only $2\times 10^{-4}$\% of the Hilbert 
space, a problem that can be solved on a personal computer. 
In the following section, we present the model and describe the 
block-spin basis. We then discuss the adaptive state selection algorithm 
employed in this work and present our results. In the final section,
we examine the strengths and weaknesses of the approach and discuss
its possible generalizations and extensions.

\section{Block-spin basis}

We wish to illustrate the method on a simple model problem, the 40-site
Heisenberg spin ${1\over 2}$ chain with periodic boundary conditions. 
The Hamiltonian is

\begin{equation}
\rm H = J \sum_{i=1}^N \Big\{ S_i^z S_{i+1}^z + {1\over 2}\big(
S_i^+ S_{i+1}^- + S_i^- S_{i+1}^+ \big)\Big\}
\label{Hamiltonian}
\end{equation}

\noindent where N=40 and $\rm S_{N+1}^\alpha \equiv S_1^\alpha$,
$\alpha = z , \pm$. In the following we set the exchange constant 
J = 1. We group the sites in blocks of 8 spins and
consider the following tensor products of block-spin states

\begin{equation}
\rm {|b^{(1)}_i>\bigotimes |b^{(2)}_j>\bigotimes |b^{(3)}_k> 
\bigotimes |b^{(4)}_l>\bigotimes |b^{(5)}_m>}
\label{Blocks}
\end{equation}

\noindent where the block states $\rm \{|b^{(m)}_p>\}$ are linear
combinations of the Ising basis states on each block:

\begin{equation}
\rm |b^{(m)}_p> = \sum_{\{\sigma\}}
C^{(p)}(\sigma_{8(m-1)},\sigma_{8(m-1)+1},\dots,\sigma_{8m})
|\sigma_{8(m-1)},\sigma_{8(m-1)+1},\dots,\sigma_{8m}>
\label{Coeff}
\end{equation}

\noindent where $\sigma_i = \uparrow,\downarrow$ is the $\rm S_z$
component of the spin on site i, 
and $\rm \{C^{(p)}\}_{p=1}^{2^8}$ are real numbers such 
that the states $\rm \{|b^{(m)}_p>\}_{p=1}^{2^8} $ form an orthonormal
basis for block m. Since the ground state is translationally invariant,
all the blocks are equivalent and thus the many-body states that
can be obtained by a cyclic permutation of the block indices will have 
the same coefficient in the wavefunction. We use this symmetry along
with the conservation of the total $\rm S_z^{tot}$ component of the
z-axis magnetization ($\rm S_z^{tot} = 0$ for the ground state) to
reduce the dimension of the matrix that must be diagonalized. Even
with these symmetries taken into account, the Hilbert space in the
ground state symmetry sector contains approximately 20 billion states.
This number is larger than the size of the matrix one would have to
diagonalize had one used the simple Ising basis of tensor products of 
single site states. The reason is that we have 
only taken into account translation invariance at the block
scale, and thus only reduces the dimension of the Hilbert space by a 
factor $\rm \sim 5$ instead of the usual factor $\rm \sim N$ when the
full translation invariance is taken into account. If one were to
construct Bloch states out of the tensor 
products, Eq.(~\ref{Blocks}), the basis
would no longer be orthonormal and the evaluation of the matrix
elements of the Hamiltonian and the calculation of the overlap
matrix would be much more complicated.

All that is required of the coefficients $\rm \{C^{(p)}\}_{p=1}^{2^8}$ 
is that they provide a orthonormal basis for a block-spin. One 
therefore has some freedom in choosing the block states, and this
should clearly be used to one's advantage. A drawback of the
block-spin construction is that it introduces artificial boundaries
between blocks in the system. The exact solution does not of course
depend on the basis, but since one has to trunctate the expansion
of the wavefunction, there will be some effect of the block 
boundaries and one ought to minimize it. To this end let us
choose the states on the blocks to be the eigenstates
of the Hamiltonian

\begin{equation}
\rm {\cal H}_{block} = \sum_{i,i+1 \in block} 
J_i \vec S_i\cdot \vec S_{i+1}
\label{BlockHam}
\end{equation}

\noindent where the coupling constants $\rm \{J_i\}$ are taken as 
parameters. The block-spin basis clearly depends on the $\rm \{J_i\}$ and
one can use the variational principle to set these parameters to values 
such that the expectation value of the full Hamiltonian with respect to
some trial wavefunction is minimized. We have found by calculating the
eigenstates of the ground state block density matrix for the 24-site chain 
that the probability of finding the block in the lowest singlet or triplet 
state is roughly 90\%. Thus we use as the trial wavefunction all of the 
possible many-body states with $\rm S^{tot}_z = 0$ that can be formed with
the lowest singlet and triplet states in each block. For the 40-site chain,
there are 52 such configurations, taking into account the equivalence 
of the blocks. We have found that taking more states of each block
into account improves the results only very slightly while increasing the
computational cost considerably. This extra computer time is much better
spent on the adaptive selection of the many-block configurations described
in the next section. Carrying out the optimization with the 52
configurations, one obtains the following values for $\rm \{J_i\}$:

\begin{equation}
\rm J_1 = J_7 = 0.33\; ;\quad J_2 = J_6 = 0.70\; ;\quad 
J_3 = J_5 = 0.81\; ;\quad J_4 = 1.00
\label{JValues}
\end{equation}

\noindent The result of this optimization is to make 
the ends of the blocks softer, allowing them to
better adapt to the presence of their neighbors. Such
smooth boundary conditions were introduced by Vekic and White\cite{Smooth} in
order to minimize finite size effects. They were also shown by the
same authors to considerably improve the "fermion sign problem"
in Quantum Monte Carlo simulations\cite{Sign}.

\section {Truncation of many-body wavefunction}

Since in practice it is not going to be possible to keep all of the
possible tensor products of block-spin states with the proper quantum
numbers, we would like to select the most important ones, namely the
many-body states with the largest weights in the ground state 
wavefunction. Consider a complete orthonormal basis $\rm \{|\phi_n>\}$.
The ground state wavefunction of the many-body system $\rm |\psi_0>$ 
can (at least in principle) be expanded as

\begin{equation}
\rm |\psi_0> = \sum_n c_n|\phi_n>
\label{Expand}
\end{equation}

\noindent We would like to truncate the expansion, Eq.(~\ref{Expand}) such
that all of the coefficients $\rm |c_n|$ are larger than some threshold
$\rm \epsilon$. The other many-body states are discarded. The results 
become exact as $\rm \epsilon \rightarrow 0$. The algorithm 
to pick the set of configurations such
that $\rm \epsilon < |c_n|$ must be adaptive, since the states
that one may wish to add to our set depend on what states we have already 
kept. For instance, there is no point in adding a state $\rm |\phi_i>$ 
with $\rm <\phi_i|H|\phi_n> = 0$ for all $\rm |\phi_n>$'s in our set, since upon 
diagonalizing the Hamiltonian in that restricted subspace, the coefficient
of $\rm |\phi_i>$ in the approximate ground state wavefunction vanishes.
 
Let us change the notation and denote $\rm \{ |\alpha> \}$ 
the set of $\rm N_s$ configurations retained (internal space), 
$\rm |\psi^{(0)}>$ the variational ground state in the 
$\rm \{|\alpha>\}$ subspace of energy $\rm E^{(0)}$, and
$\rm \{ |\beta> \}$ the set of states not retained (external space).
We first estimate the correction $\rm |\delta\psi>$ from the external 
space to the approximate eigenstate $\rm |\psi^{(0)}>$ by
first order perturbation theory, using as unperturbed Hamiltonian

\begin{equation}
\rm {\tilde H}_0 = |\psi^{(0)}>E^{(0)}<\psi^{(0)}|
\label{H0}
\end{equation}

\noindent and

\begin{equation}
\rm {\tilde H}_1 = H - {\tilde H}_0 
\label{Hint}
\end{equation}

\noindent as the perturbation. One thus has

\begin{equation}
\rm |\psi^{(0)}> \equiv \sum_{\alpha} c_{\alpha} |\alpha>
\label{psi0}
\end{equation}

\begin{equation}
\rm |\delta\psi> \equiv \sum_{\beta} c_{\beta} |\beta>
\label{dpsi0}
\end{equation}

\noindent where the coefficients $\rm \{ c_{\alpha} \}$ where obtained
by diagonalizing the Hamiltonian in the internal space and we wish to
estimate the coefficients $\rm \{ c_{\beta} \}$. To first order in
$\rm {\tilde H}_1$, one has

\begin{equation}
\rm c_{\beta} = {1\over E^{(0)}} \sum_{\alpha} 
c_{\alpha} <\alpha|H|\beta>
\label{cbeta}
\end{equation}

\noindent At this stage, we discard the configurations for which 
$\rm |c_{\beta}| < \epsilon' < \epsilon$ (in practice 
$\rm \epsilon' = 0.5\epsilon$ works fine). The rationale for estimating
the correction to $\rm |\psi^{(0)}>$ by perturbation theory first is 
that one would like to identify the configurations that have weights
much smaller than the threshold and eliminate those right away. This
way the internal space does not expand too much at each iteration,
the dimension of the matrix that one must diagonalize is significantly
smaller and it saves a significant amount of computer time. On the other hand,
since we only estimate the coefficients to first order, we are making 
some error and in the event the coefficients of some states are 
underestimated in the calculation, one might throw away states that 
should be kept. The present method compromises by keeping states
with coefficients slightly smaller in magnitude than the final cutoff.
It is found in practice that the error introduced by this early
truncation is order of magnitudes smaller than the difference between
the variational energy and the exact result (typically in the sixth digit).
It is therefore perfectly legitimate and speeds up the computation by
at least one order of magnitude. Next, we diagonalize the Hamiltonian
matrix in the expanded internal space and the states whose coefficients
$\rm c_{\alpha}$ are smaller in magnitude than the threshold 
$\rm \epsilon$ are discarded. The procedure can be repeated a number of
times, until no more states whose coefficients are larger in magnitude
than $\rm \epsilon$ can be added to the set.

This method works very well for any type of basis. However, there is
one feature of the block-spin basis that requires a couple more tricks. In the
basis that consists of tensor products of single spin states (Ising basis), 
the Hamiltonian acting on a state can produce at most N (the number of sites)
other states, since there are N spin-flip 
terms in the Hamiltonian, Eq.(\ref{Hamiltonian}).
Since the total number of states grows exponentially with N, the matrix
representation of the Hamiltonian in this simple basis is extremely sparse. 
The block-spin states are linear combinations of many Ising basis states 
(up to 70 in the $\rm S_z=0$ sector of the eight spin block) and thus each 
member of the many-body basis of tensor products of block-spin states 
contains a very large number of Ising basis states.
The matrix representation of the Hamiltonian in this block-spin basis
is not very sparse and the application of the Hamiltonian operator to any
internal state $\rm |\alpha>$ produces a very large number of 
external states $\rm |\beta>$. If the Hamiltonian matrix were full,
one could generate the entire external space by applying the Hamiltonian
to a single state. If the matrix is not completely full, there is still
some redundancy and it is not necessary 
to apply the Hamiltonian to every state to 
generate the external space (of course in general we do not know which 
subset of states will generate the complete external space). We also expect 
that the next most important states are those that directly couple 
to the most important states we already have
(that actually depends on the quality of the basis, i.e. the smallness of
off-diagonal matrix elements of the Hamiltonian for states far 
apart in energy). We have used these heuristics to modify the algorithm
as follows: one generates the states $\rm |\beta>$
in the external space by applying the Hamiltonian only to the most
important states in the internal space, namely those with coefficients
larger in magnitude than $\rm \epsilon" > \epsilon$. 
In practice, one must test various values of 
$\rm \epsilon"$ to determine a good compromise between accuracy 
and speed of computation. We have found that it is possible to choose
$\epsilon"$ such that the error introduced is much smaller than the error
due to the truncation of the Hilbert space. Another simplification
is to restrict a priori the number of block states that are to be
used. In the early steps of the algorithm, there is no need to
consider very highly excited states of the blocks. This restriction can
be loosened as the adaptive state selection proceeds. One can also do
that in such a way that the approximation introduces an error less
than the systematic error due to the truncation of the many-body 
wavefunction. These tricks speed up the algorithm by at least an 
order of magnitude. It is of course possible to relax each of these
approximations to estimate the effect they have on the final 
results.

By studying the increase in the number of states kept and 
the distribution of the weights of these states 
as a function of their energy in the variational
ground state wavefunction as the cutoff $\rm \epsilon$ is decreased,
the adaptive state selection algorithm described above can give us a 
good idea of the quality of the basis used and how the off-diagonal
matrix elements decrease as the energy difference between states
increases.

\section{Results}

It is important that the method not only give accurate results for
local quantities such as the ground state energy, but it
should also yield accurate results for the long-range spin 
correlations. The ground state spin-spin correlation function
$\rm C(l) = <\vec S(i)\cdot\vec S(i+l)>$ is shown in Fig.~\ref{fig1}
where it is compared to the Quantum Monte Carlo results
of Sandvik\cite{Sandvik}. Since the method only takes into account
the equivalence between blocks and not the full translation 
invariance, the results of this calculation would actually 
depend on the site i. The quantity C(l) shown in Fig.~\ref{fig1} is
averaged over i: 

\begin{equation}
\rm C(l) = {1\over N}\sum_i <\vec S(i)\cdot\vec S(i+l)>
\label{cfunc}
\end{equation}

\noindent In this calculation, the adaptive state selection algorithm
described in the previous section was used with a cutoff
$\rm \epsilon = 5\times 10^{-4}$. The variational ground state
wavefunction obtained is a linear superposition of 8553 (out of
a possible 20 billion) basis states
(therefore about $2\times 10^{-4}\%$ of the Hilbert space),
and only 56 out of the 256 block basis states were actually used.
The other parameters used in the calculation were 
$\rm \epsilon' = 0.5\epsilon$ and $\rm \epsilon" = 5\times 10^{-3}$.
The variational ground state energy obtained is -17.728 J compared to 
the exact value of -17.7465227883 J from the Bethe ansatz\cite{Ha}.
Given the energy of the mean-field (N\'eel) state of -10 J, one 
recovers 99.8\% of the correlation energy. One does a good job with
the long-range spin correlations. Note that for a separation of 16
lattice sites and over, the spins are not even in nearest neighbor
blocks. The error with respect to the Monte-Carlo results is about
1.5\% for C(l=20). It is instructive to look at what the spin-spin
correlations look like in the block-spin mean-field ground state

\begin{equation}
\rm |\Psi_{MF}> =
|b^{(1)}_0>\bigotimes |b^{(2)}_0>\bigotimes |b^{(3)}_0> 
\bigotimes |b^{(4)}_0>\bigotimes |b^{(5)}_0>
\label{BlockMF}
\end{equation}

\noindent with the smooth boundary conditions and with open
boundary conditions. In the above equation
$\rm |b^{(i)}_0>$ is the ground state of block i.
The results are shown in Fig.~\ref{fig2}. For nearest neighbor spins 
(i.e. for the energy), the block-spin states with open boundary
conditions give a lower variational energy. On the other hand,
for correlations at larger distances, the
block-spin states with the smooth boundary conditions do a much
better job. Note that a single tensor product of block-spin states
does not correlate spins in different blocks and thus the correlation
function C(l) = 0 when $\rm l \geq 8$, as it should.
If one uses the block-spin basis with open boundary
conditions, one needs nearly ten times as many basis functions
(approximately 80,000) to obtain results of quality similar to 
those shown in Fig.~\ref{fig1}. It is thus worthwile to slightly
increase the ground state energy of the block in order to
better describe the spin-spin correlations at distances 
larger than one lattice spacing.

We have also looked at choosing the basis on a block in the spirit
of the DMRG, namely by calculating the block density matrix from
some approximate ground state wavefunction and using as the block
basis the eigenstates of the block density matrix. One can of
course iterate the process and recompute an approximate ground
state wavefunction using the improved block basis, and using the
eigenstates of the new block density matrix. The process converges
in a few iterations. It turns out, however, that the 
block basis obtained in this way is very similar to the one one
gets with the smooth boundary conditions, and leads to results
of the same quality.

We have mentioned in the introduction that one would like to have
a basis such that the importance of the states decreases as their
energy increases. In Fig.~\ref{fig3} we show the absolute value of the
coefficients of the 8553 basis states in the variational wavefunction 
used in the calculation of C(l) shown in Fig.~\ref{fig1} as a function
of their energy. At low energy, the weights of the states decay
roughly exponentially as a function of energy, but the situation
totally deteriorates at higher energies, where there is no longer
a clear hierarchy of the basis states. We speculate that this is
due to the effect of the block boundaries. The basis states do not
have all of the symmetries of the exact ground state, and this
ultimately limits the accuracy achievable with this method. The
smooth boundary conditions do not completely eliminate the problem.
In order to obtain much more accurate results than those shown in 
Fig.~\ref{fig1}, one would have to retain an exponentially 
larger number of states. As one can see
from Fig.~\ref{fig3}, for cutoffs $\rm \epsilon$ much
smaller than $5 \times 10^{-4}$, one would have to keep states
in a wide energy range and since the density of states increases 
exponentially with energy, one needs to keep an exponentially 
larger number of states.

\section{Outlook}

We have shown that an expansion of the ground state many-body 
wavefunction in terms of a block-spin basis allows a severe 
truncation of the Hilbert space while maintaining good 
accuracy for a 40-site Heisenberg spin-1/2 chain. For this 
problem, the number of basis states one must use (for a given
accuracy) grows exponentially with the number 
of blocks in the system: spins in any two 
blocks, however far apart, must be correlated.
Model Hamiltonians whose excitation spectrum possesses a gap in the 
thermodynamic limit do not of course present this difficulty.
In the present case, however, this ultimately limits the size of
the system that can be treated accurately. For a spin-1/2 system,
it is not inconceivable to use blocks of 12 spins, and with five such
blocks one should be able to tackle systems with up to 60 spins,
keeping on the order of 10,000-20,000 many-block configurations.
We have not attempted this, since it would have required an extensive 
reprogramming effort due to memory constraints on the computer system
used for these calculations. Note that the DMRG method for this 
problem is superior, in that much larger systems
can be treated with ease. The purpose of the present work
was not an attempt to supersede the DMRG in one dimension 
(in the opinion of the author, this extremely unlikely to happen),
but to study an alternative many-body basis in the hope that
it could be used when the DMRG method is not as effective.

The block-spin approach is trivially extended to two dimensions.
We have in fact carried out calculations for two-dimensional spin-1/2
Heisenberg systems. In higher dimensions, the quantum fluctuations 
are weaker than for the one-dimensional spin chain, and as a result 
less correlation energy must be recovered given a fixed number of
spins. We therefore expected the two-dimensional problem to be
easier to solve with the present technique. Unfortunately,
it has turned out otherwise. The possibility of long-range order
in two dimensions is at the root of the problem. Since the
ground state of the two-dimensional Heisenberg model possesses 
long-range order, the correlations of spins in widely separated blocks
are such that they saturate to a finite constant value
as the separation between the blocks goes to infinity. On the other
hand, our block-spin basis is such that spins in different blocks
are completely uncorrelated. In one dimension, one can build up these
block-block correlations with some effort, since they decay algebraically.
Uncorrelated spins a distance 8 apart provides quite a good
starting point. In two-dimensions, one has to start with uncorrelated
spins a distance 3 or 4 apart ($3 \times 3$ or $4 \times 4$ are clearly the 
maximum block sizes that can be handled) and it turns out to be a much
poorer starting point. The possibility of long-range order in 
two-dimensions must therefore be incorporated in the basis in some
way. We plan to look at various possibilities.

As the DMRG algorithm has demonstrated, computational methods
based on an approximate diagonalization of the Hamiltonian can be
competitive with Monte Carlo methods. The success of the DMRG
and the results presented in this paper indicate that such an
approximate diagonalization is only effective if carried out 
in a correlated basis that closely approximates the true
many-body ground state correlations. The challenge is to
find such a basis for the many two and three-dimensional
model Hamiltonians which remain unsolved.

I should like to thank Z. Ha, E. Manousakis, and A. Sandvik for
discussions on these and related topics. This work was supported
by the National High Magnetic Field Laboratory and the State of
Florida.

\begin{figure}
\caption{The spin-spin correlation function, Eq.~(\protect\ref{cfunc}), 
with the adaptive state selection parameters $\epsilon = 5\times 10^{-4}$, 
$\epsilon' = \epsilon/2$ and $\epsilon" = 5\times 10^{-3}$ (diamonds) is
compared to the Monte Carlo results of Sandvik\protect\cite{Sandvik} 
(dashed line and error bars).
\label{fig1}}
\end{figure}

\begin{figure}
\caption{The spin-spin correlation function, Eq.~(\protect\ref{cfunc}), 
for the block mean-field state, Eq. ~(\protect\ref{BlockMF}), 
is shown with smooth boundary conditions on the blocks
(diamonds) and with open boundary conditions on the blocks
(crosses).
\label{fig2}}
\end{figure}

\begin{figure}
\caption{Writing the variational ground state 
wavefunction $\rm |\psi_0> = \sum_i c_0(i)|\phi_i>$ 
in the basis $\rm \{|\phi_i>\}$, we show the magnitude of
the coefficients $\rm c_0(i)$ for the block-spin
basis state $\rm |\phi_i>$ as a function of its energy 
$\rm <\phi_i|H|\phi_i>$. The variational wavefunction used is obtained
from the adaptive state selection algorithm with the same parameters
as described in Fig.~\protect\ref{fig1} and in the text.
\label{fig3}}
\end{figure}

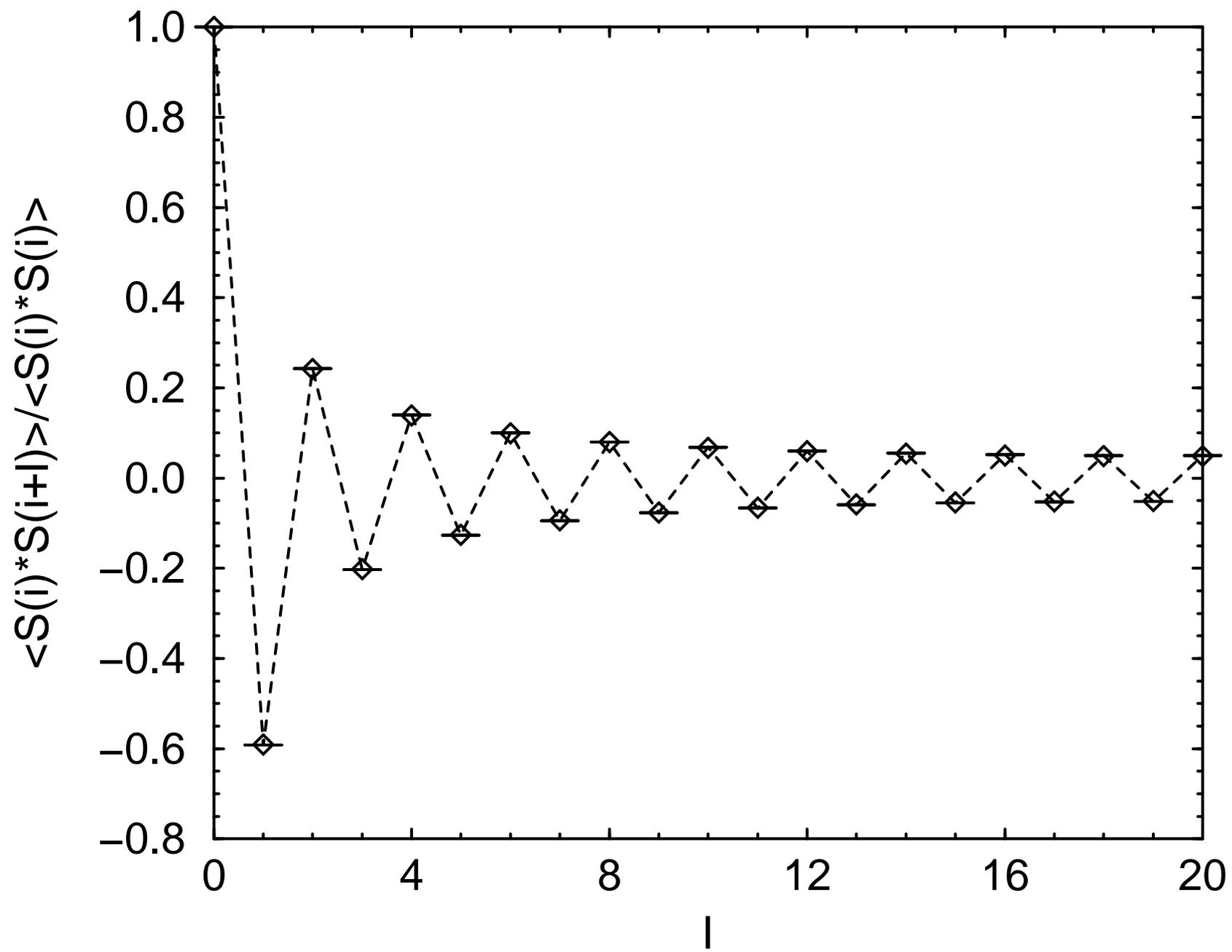
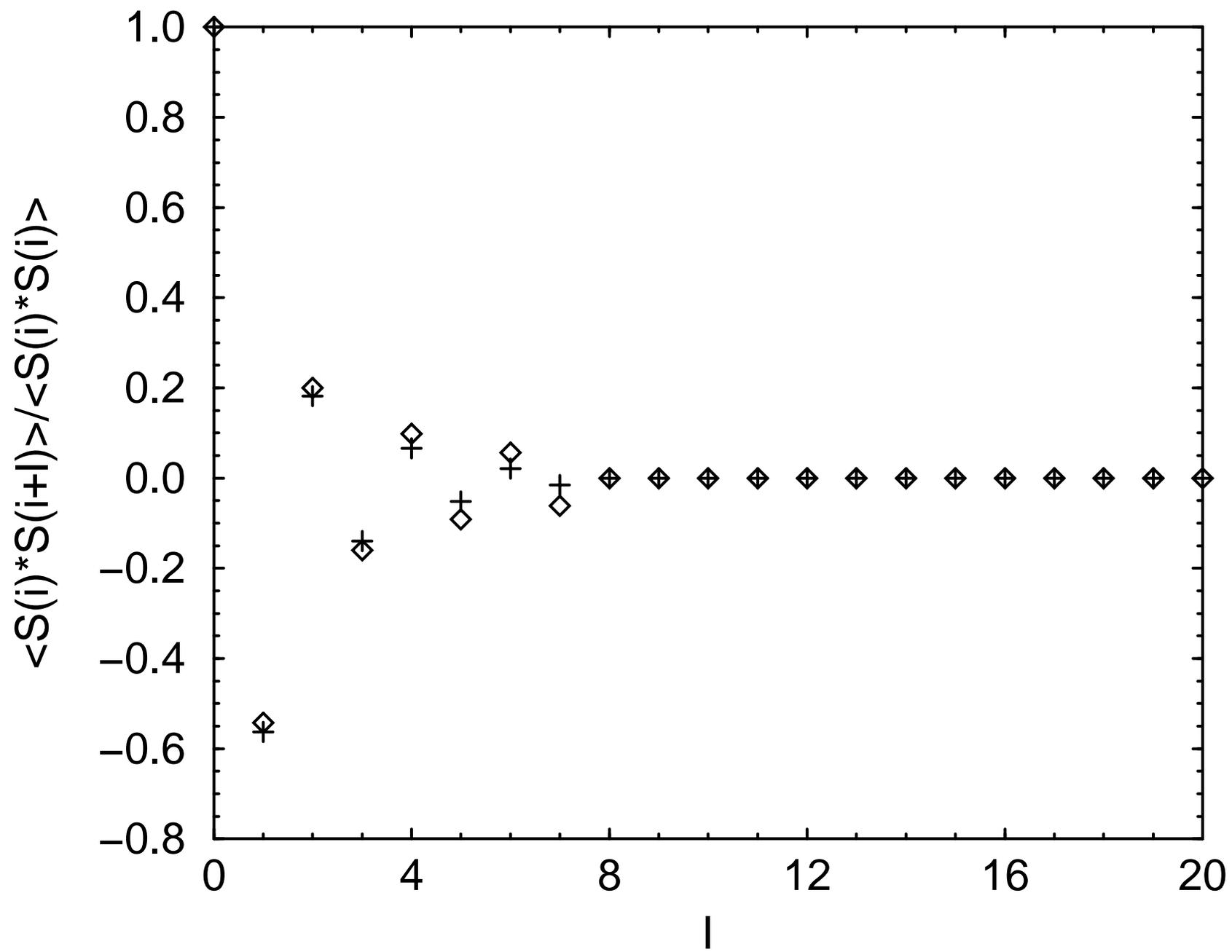
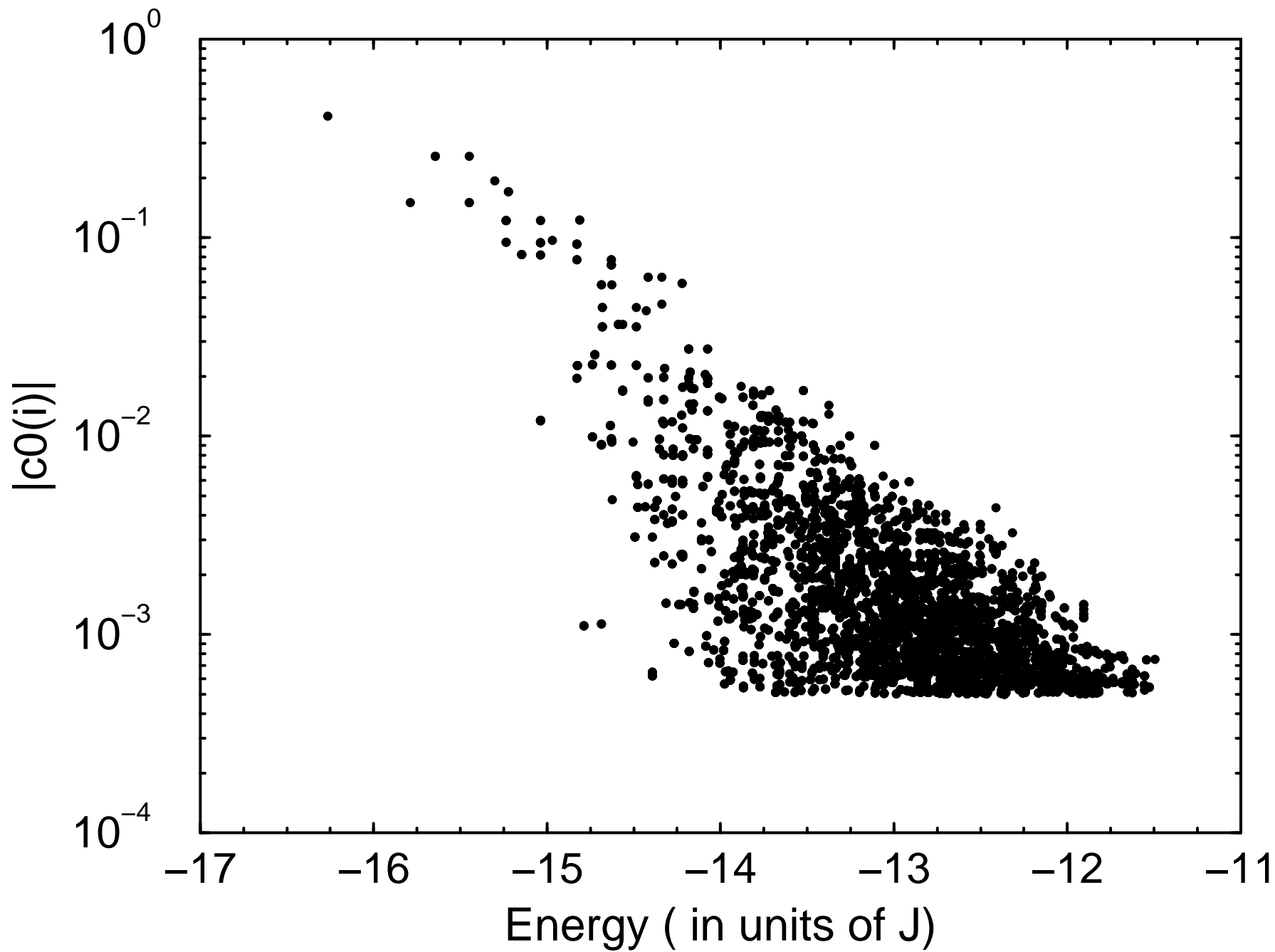
\end{document}